\documentclass{article}
\usepackage[letterpaper, portrait, margin=1.5in]{geometry}
\usepackage[utf8]{inputenc}
\usepackage[sort&compress, square]{natbib}
\setcitestyle{numbers}
\usepackage{xurl}

\begin{document}

\title{Report of the Topical Group on Quantum Sensors for Snowmass 2021}

\author{Thomas~Cecil, Kent~Irwin, Reina~Maruyama, Matt~Pyle, Silvia~Zorzetti\\ \\ \small{Contributors from the community}:\\\small Asher Berlin, Sergey Belomestnykh, Diego Blas, Anthony J. Brady,\\\small Caterina Braggio, Oliver Buchmuller, Dmitry Budker, Marcela Carena,\\\small Daniel Carney, Raphael Cervantes, Mattia Checchin, Timothy E. Chupp,\\\small Crispin Contreras-Martinez, Raffaele Tito D’Agnolo, John Ellis,\\\small Sebastian A. R. Ellis, Grigory Eremeev, Daniil Frolov, Andrew A. Geraci,\\\small Christina Gao, Bianca Giaccone, Anna Grassellino, David Hanneke,\\\small Roni Harnik, Jason Hogan, Matthew Hollister, Nicholas R. Hutzler,\\\small Ryan Janish, Andrew Jayich, Yonatan Kahn, Sergey Kazakov,\\\small Derek F. Jackson Kimball, Shimon Kolkowitz, Zhen Liu, Andrei Lunin,\\\small Doga Murat Kurkcuoglu, Oleksandr Melnychuk, Gavin W. Morley,\\\small Holger Muller, Alexander Netepenko, Zachary Pagel, Cristian Panda,\\\small Roman Pilipenko, Yuriy Pischalnikov, Sam Posen, Surjeet Rajendran,\\\small Alex Romanenko, R.F. Garcia Ruiz, Marianna S. Safronova,\\\small Jan Schutte-Engel, Jaideep T. Singh, Alexander O. Sushkov,\\\small Changqing Wang, Vyacheslav Yakovlev, Kevin Zhou, and Quntao Zhuang}

\date{August 2022}

\maketitle

\abstract{Quantum Sensors offer great potential for providing enhanced sensitivity in high energy physics experiments. In this report we provide a summary of key quantum sensors technologies - interferometers, optomechanics, and clocks; spin dependent sensors; superconducting sensors; and quantum calorimeters - highlighting existing experiments along with areas for development. We also provide a set of key messages intended to further advance the state of quantum sensors used for high energy physics specific applications.}


\section{Executive Summary}
The use of quantum sensors in high energy physics has seen explosive growth since the previous Snowmass Community Study. This growth extends far beyond high energy physics (HEP) impacting many areas of science from communications to cryptography to computing.  Quantum sensors have been used in searches for dark matter - particle and wave, fifth forces, dark photons, permanent electric dipole moment (EDM), variations in fundamental constants, and gravitational waves, among others. These sensors come in a wide range of technologies: atom interferometers and atomic clocks, magnetometers, quantum calorimeters and superconducting sensors to name a few. Early work with quantum sensors in the context of particle physics often focused in cosmic and rare and precision frontiers, but recent concepts seek to expand the use of quantum sensors to the energy and neutrino frontiers solidifying them as fundamental technologies for the future of experimental HEP. Based upon input to the Snowmass process our topical group has identified several key messages necessary to support the development and use of quantum sensors in HEP:




 \begin{itemize}
    \item \textbf{Continue strong support for a broad range of quantum sensors. Quantum sensors address scientific needs across several frontiers and different technologies carve out unique parameter spaces.} While these sensors share many common characteristics, each has advantages that make it the sensor of choice for specific applications along with challenges that need further development to make the greatest impact.  
    \item \textbf{Continue support for R\&D and operation of table-top scale experiments. Many are shovel ready and have the potential for large impact.} Much of the growth in quantum sensors over the past decade has occurred in small, laboratory based experiments. 
    These fast-paced small experiments should continue to be supported as a way to rapidly develop sensor technology and help determine those areas where quantum sensors can have the greatest impact.
    \item \textbf{Balance support of tabletop experiments with pathfinders R\&D to address the large-scale challenges of scaling up experiments which will require National Lab and HEP core competences.} As the fast-paced, small experiments mature, those with significant discovery potential begin to emerge along with areas of commonality between the experiments (e.g.\ the need for advanced high field magnets for axion dark matter experiments or ultra-stable lasers for atom interferometers and clocks). They have reached the point at which plans for larger-scale, longer-term experiments should be conceptualized. These concepts can evaluate the potential reach that can be achieved in a larger effort and the scale of require technological development. 
    \item  \textbf{Develop mechanisms to support interactions outside of the HEP program to enable collaborations with fields with developed expertise in quantum sensors. Advances in quantum information science (QIS) provide exceptional theoretical and experimental resources to advance quantum sensing that could provide mutual benefits in several areas such as materials, detectors, and devices.} Many of the most promising quantum sensors for HEP science have been developing for the past decade or more in areas outside of the traditional HEP science and funding sphere. For example, atomic clocks developed over many decades as a source of precision timing standards are now stable enough they can be used in the search for variations of fundamental constants and gravitational waves. The HEP community should strive to collaborate with these broader communities in a way the gives HEP access to new sensor technologies while sharing HEP expertise (e.g.\ large magnets and vacuum systems). Effort should be made to allow the free flow of ideas and effort across traditional funding boundaries to encourage scientists and engineers working with quantum sensors to tackle the most interesting and challenging problems available. 
    \item \textbf{Develop mechanisms to facilitate interactions to support theoretical work address issues of materials and measurement methods.} As with other instrumentation frontiers and as quantum sensors become more sensitive, focused support on quantum materials at the interface of quantum sensors and HEP will be needed. This includes theoretical work necessary for on topics including quantum materials, squeezing, and back action.
    \item \textbf{Workforce development is needed to encourage workers with the needed skills to engage with the HEP field, maintain current momentum, and for long-term success in the face of growing competition from industrial quantum computing.} While the high energy physics community is poised to benefit from quantum sensor developments outside of HEP, we face a shortage of skilled workers. The explosive growth in quantum computing in recent years, along with arrival of several major tech companies has created a fierce competition for workers with skills needed to develop quantum sensors and experiments. The HEP community will need to invest now in order to train and retain the next generation of quantum scientist. Increasing collaborations outside of HEP -- as discussed above -- can provide an additional pathway to reaching skilled workers and engaging them on HEP challenges.  
    
\end{itemize}

\section{Overview}

\subsection{Introduction}
In this report we provide an overview of recent development in quantum sensors and their scientific impact to the high energy physics community as presented in the many Letters of Intent (LOIs) and white papers submitted during the Snowmass 2021 process. We focus on sensors in which the quantum state of the sensor can be measured and manipulated. Also included are quantum calorimeters to measure individual quanta of energy deposited in the sensor. As a group, quantum sensors are extremely sensitive devices used to explore new physics. The goal is to use hardware and manipulation techniques developed in quantum information science and technology to reach sensitivities better than the standard quantum limit (SQL) over as broad a bandwidth as possible. 

Many of the most promising quantum sensors for HEP science have been developed over the past decade or more in areas outside of the traditional HEP science and funding sphere. For example, atomic clocks, developed over many decades as a source of precision timing standards, are now stable enough they can be used in the search for variations of fundamental constants and gravitational waves. At the same time, the field of quantum computing is experiencing multiple breakthroughs. We point the reader to the report from the Quantum Computing topical group in the Computational Frontier for more details, but highlight that many of the technological developments needed to improve quantum computers are also needed to improve quantum sensors: longer coherence times, increases numbers of quantum states, isolation from environmental noise, etc. Quantum sensing and computing also both require a highly skilled workforce. With the arrival of several major tech companies in the field of quantum computing there is fierce competition for workers. The HEP community will need to invest now in order to train the next generation of quantum scientist. 

Much of the growth in quantum sensors over the past decade has occurred in small, laboratory based experiments. These small experiments have allowed the broader community to try out many different varieties of sensors aimed at a broad range of scientific targets (see Sec.~\ref{Technologies} for examples). The extreme sensitivity of quantum sensors and quantum techniques often enables these experiments to make significant gains in unexplored parameter spaces at minimal cost while also advancing the sensor technology. Continued support of these `table top' experiments serves as a way to rapidly develop sensor technology and help determine those areas where quantum sensors can have the greatest impact.
As the fast-paced small experiments mature, those with the significant discover potential begin to emerge along with areas of commonality between the experiments (e.g.\ the need for advanced high field magnets for axion dark matter experiments or ultra-stable lasers for atom interferometers and clocks). They have reached the point at which plans for larger-scale, longer-term experiments are being developed. These concepts can evaluate the potential reach that can be achieved in a larger effort and the scale of require technological development.

\subsection {Science}
Quantum sensors encompasses a broad spectrum of technologies (as described in Section \ref{Technologies}) and have the potential to impact a wide range of core HEP science. This can seen in the broad collection of white papers submitted to this community study that make use of these sensors: dark matter - axion, wavelike, and particle from ultra-light to ultra-heavy \cite{Jaeckel.2022,Antypas.2022,Carney.20223m,CDMS.2022,Ebadi.2022,Collaboration.2022,Wang.2022,Essig.2022}, new particles or forces \cite{Berlin.2022}, the electric dipole moment (EDM)\cite{Alarcon.2022}, variations in fundamental constants, gravitational-wave detector facilities \cite{Ballmer.2022}, spacetime symmetries \cite{Adelberger2022}, and neutrino masses \cite{Armatol.2022, Ullom.2022}.

\section{Technologies} \label{Technologies}
\subsection{Interferometers, Optomechanics, and Clocks}
\emph{Note : The contents of this section are in part taken and modified from: Snowmass 2021: Quantum Sensors for HEP Science -- Interferometers, Mechanics, Traps, and Clocks \cite{Carney.20223m}}

\subsubsection*{Atom Interferometers}
Atom interferometry is a growing field with a variety of fundamental physics applications including gravitational wave detection, searches for ultralight (wave-like) dark matter candidates and for dark energy, tests of gravity and searches for new fundamental interactions (“fifth forces”), precise tests of the Standard Model  (e.g. fine structure constant), and tests of quantum mechanics. In light-pulse atom interferometry, laser pulses are used to coherently split, redirect, and recombine matter waves. 
Conventional atom interferometry makes use of a pair of counter-propagating laser beams to drive two-photon Raman or Bragg transitions while a new variation takes advantage of long-lived excited states in alkaline-earth-like atoms that can be resonantly driven by a single laser beam. In a gradiometer configuration, two identical atom interferometers are run simultaneously on opposite ends of a baseline, using the same laser sources. A comparison of the individual atom interferometer signals yields a differential measurement that enables the cancellation of noise common to both interferometers. This in principle enables superior common-mode rejection of noise, allowing for the possibility of, for example, gravitational wave detection using a single baseline. A passing gravitational wave would modulate the baseline length, while coupling to an ultralight dark matter field can cause a modulation in the energy levels. This combines the prospects for both gravitational wave detection and dark matter searches into a single detector design, and both science signals are measured concurrently.


As one example, the MAGIS concept takes advantage of features of both clocks and atom interferometers to allow for a single-baseline gravitational wave detector. MAGIS-100 is the first detector facility in a family of proposed experiments based on the MAGIS concept. The instrument features a 100-meter vertical baseline and is now under construction at the Fermi National Accelerator Laboratory (Fermilab). State-of-the-art atom interferometers are currently operating at the 10-meter scale, while a kilometer-scale detector is likely required to detect gravitational waves from known sources. The Atom Interferometric Observatory and Network (AION) project envisages a staged Atom Interferometry program, starting with a 10 m device and progressing via a 100 m experiment to a 1 km instrument. AION will enable exploration of the properties of ultra-light dark matter (DM) and gravitational waves (GWs) from the very early Universe and astrophysical sources in the mid-frequency band ranging from several mHz to a few Hz, intermediate between the sensitive ranges of LIGO/Virgo/KAGRA and LISA. The ultimate sensitivity of the AION program will be reached by interoperating and networking with other instruments around the world, similar to the existing LIGO-Virgo network, which will provide science opportunities not accessible to single detectors.

\subsubsection*{Optomechanical Sensors}
Mechanical sensors that can be read read out optically (frequencies range from microwave to visible) have advanced rapidly and are now commonly operated in a regime where their sensitivity becomes dominated by quantum noise in the mechanics or readout system. A wide variety of sensors is available, ranging from single ions to kilograms-scale elements. The sensors are uniquely suited to coherent signals with a scale comparable to the size of the mechanical sensor since the signal is coherently integrated into a small number of degrees of freedom (e.g.\ center of mass motion). A key example of the capabilities of optomechanical sensors is LIGO. Other examples include mechanically suspended reflective pendula; optically levitated dielectrics, cold atoms, and ions; clamped nanomechanical membranes; magnetically levitated systems: and can also include collectively quantized degrees of freedom like phonons. 

In addition to their use in gravitational wave detection and precision measurements in metrology, optomechanical devices are rapidly being incorporated into the portfolio of detector systems useful for a number of high energy and particle physics targets. Building on classical proposals for neutrino and dark matter detection with nanoscale targets, proposals now exist to use optomechanical sensors for detection of ultra-light, MeV-to-TeV scale, and ultra-heavy dark matter \cite{Windchime.2022}); neutrinos; high-frequency gravitational waves; fifth-force modifications to Newton’s law at tabletop scales; deviations from standard quantum mechanics (including ideas about gravitational breakdown of quantum mechanics); and tests of quantum properties of the gravitational interaction. 

Moving forward, a number of key opportunities exist to increase the utility of these devices in the search for new physics. There is a critical need for new theoretical ideas about potential new signals. There is a push to improve detector technologies to reach sensitivities at and beyond the so-called Standard Quantum Limit (SQL). The most common, well-demonstrated method to go beyond the SQL is the use of squeezed light while a less-studied, but enticing option is the use of back action evasion techniques. Further theoretical development and implementation of these techniques in disparate situations and physical architectures, especially in broadband sensing problems, will be of crucial importance in the next decade. Leveraging multiple sensors (“networks”) and entanglement between them can similarly enable detection beyond the SQL; using these ideas in searches for new physics would be extremely interesting.

\subsubsection*{Clocks and Precision Spectroscopy for Particle Physics}
Optical clock precision has improved by more than three orders of magnitude in the past fifteen years, enabling tests of the constancy of the fundamental constants and local position invariance, dark matter searches, tests of the Lorentz invariance, and tests of general relativity. All current atomic clocks are based on either transitions between the hyperfine substates of the atomic ground state (microwave clocks) or transitions between different electronic levels (optical clocks). The frequency ratio of two optical clock frequency is only sensitive to the variation of the fine structure constant $\alpha$ and optical atomic clocks can probe the standard matter -- dark matter coupling. Promising searches for ultralight particles are feasible through isotope-shift atomic spectroscopy, which is sensitive to a hypothetical fifth force between the neutrons of the nucleus and the electrons of the shell. The analysis of precision isotope shift (IS) spectroscopy sets limits on spin-independent interactions that could be mediated by a new particle which could be associated with dark matter. Deployment of high-precision clocks in space could open the door to new applications, including precision tests of gravity and relativity, searches for a dark-matter halo bound to the Sun, and gravitational wave detection in wavelength ranges inaccessible on Earth. Space-based optical lattice atomic clocks could potentially include the possibility of a tunable, narrowband GW detector that could lock onto and track specific GW signals provide a compliment to other experiments (e.g.\ LISA and LIGO). Radioactive atoms and molecules offer extreme nuclear nuclear charge, mass, and deformations, and may be worked with efficiently with the advanced quantum control toolset of AMO. These rare systems offer an unprecedented amplification of both parity- and time-reversal violating properties.
 
Several potential pathways existing for improving clock performance: developing new clocks with much larger sensitivity factors; development of large and more integrated clock networks (e.g.\ QSNET \cite{Barontini.2022}); making clocks more portable (critical for space applications); and improving local oscillator technology as it limits coherent integration times. Additionally it is possible to probe multiple clocks with the same laser  to cancel out local oscillator noise (similar to using single laser with atom interferometer). This pushes to near SQL. Pushing beyond the SQL can be achieved by using entangled states, such as spin-squeezed states. Gains can also be made by moving to clocks using highly charged ions (also a promising avenue for isotope shift spectroscopy) or nuclear clocks which have much higher sensitivities to the variation of alpha, up to 4 orders of magnitude for nuclear clocks. Nuclear clocks are highly sensitive to the hadronic sector and could offer improvements in sensing of DM coupling by 5-6 orders of magnitude. Use of molecular clocks provide direct sensitivity to determining the ratio of the proton mass to electron mass and its variation. 

\subsection{Spin Dependent Sensors}
\emph{Note : The contents of this section are in part taken and modified from: Quantum Sensors for High Precision Measurements of Spin-dependent Interactions. Arxiv (2022). \cite{Budker.2022}}

Experimental techniques for precision measurement of spin-dependent interactions have substantially advanced over recent decades, in no small part because control and measurement of spins, spin ensembles, and quantum materials is at the heart of QIS and quantum computing and they share a common foundation with the robust program of research on spin-based quantum sensors for measurement of magnetic fields, magnetic resonance phenomena, and related phenomena. There are three main ways measurements of spins can probe for new physics: new physics can break symmetries of the Standard Model giving rise to novel responses of spins to other fields (e.g. searching for  EDM); the new physics can directly affect the spin for example via an interaction with a new field and the spin (e.g. searches for axions and axion-like particles); and new physics can affect the environment of the spin which the spin can sense (e.g. damage to crystals containing defects centers following interaction new physics such as dark matter particles).

\subsubsection*{Electric Dipole Moments}
The general approach of electric dipole moment (EDM) experiments is to search for the combined effect of a P- and T-odd Hamiltonian and an applied electric field E, which results in an energy shift for a given quantum state of the atom or molecule. Typically the system is spin polarized via optical pumping or some other hyperpolarization technique such that the system is an a superposition of quantum states with opposite EDM-induced energy shifts. Thus a nonzero EDM will cause the polarized spins to precess in the presence of an electric field. There are several general areas of technology development that can advance the fundamental sensitivity of EDM searches: increase the energy shift by finding system with maximum enhancement factors; improve control techniques to increase the total number of polarized atoms/molecules, and achieving longer spin-coherence times.

\subsubsection*{Magnetometers}
Many theories predict the existence of new force-mediating bosons that couple to the spins of Standard Model particles. One of the primary experimental strategies is to employ a sensitive detector of torques on spins and then bring that spin-based torque sensor within a Compton wavelength  of an object that acts as a local source of an exotic field (e.g., a large mass or highly polarized spin sample). Since the observable in these experiments is a spin-dependent energy shift a sensor employing N independent spins with coherence time $\tau$ has a shot-noise-limited sensitivity. Common sensors include NV centers, optical atomic magnetometers and Bose-Einstein condensates (BEC). One promising technology is the development of levitated ferromagnetic torque sensors (LeFTS). The active sensing element consists of a hard ferromagnet, well isolated from the environment by, for example, levitation over a superconductor via the Meissner effect. The mechanical response of the levitated ferromagnet to an exotic spin-dependent interaction can be precisely measured using a superconducting quantum interference device (SQUID). Similar to the LeFTS concept, ultracold twobody interactions in the BEC create a fully coherent, single-domain state of the atomic spins that enables the system to evade the sensitivity limits of traditional spin-based sensors. 

Beyond the intrinsic sensitivity, the principal challenge in experiments searching for exotic spin-dependent interactions is understanding and eliminating systematic errors: clearly distinguishing exotic spin-dependent interactions from mundane effects due to, for example, magnetic interactions. By comparing the response of two different systems, effects from magnetic fields can be distinguished from effects due to exotic spin-dependent interactions. This is the essence of comagnetometry, where the same field is simultaneously measured using two different ensembles of atomic or nuclear spins. This effort can be extend to searching for transient interactions through the use of networks of geographically distributed spin-dependent sensors. For example the GNOME network will search for transient and stochiastic effects that could arise from ALP fields of astronomical origin passing through the Earth.

\subsubsection*{Magnetic Resonance}
One possible manifestation of ultralight bosonic dark matter is as classical fields oscillating at the Compton frequency. The bosonic dark matter field can cause spin precession via couplings to nuclear and electron spins  which can be detected using the broad and versatile tools of magnetic resonance. In a dark matter haloscope experiments, the oscillating field is assumed to always be present, corresponding to case of continuous-wave NMR. The magnetic field is scanned, and if the Larmor frequency matches the Compton frequency, a resonance occurs generating a time-dependent magnetization that can be measured, for example, by induction through a pick-up loop or with a SQUID. This is the method used in the CASPEr experiment. A key to CASPEr’s sensitivity is the coherent “amplification” of the effects of the axion dark matter field through a large number of polarized nuclear spins. Therefore an important technological development is the ability to carry out NMR on the largest possible number of spins.

The QUAX (QUaerere AXion) experiment  searches for axion dark matter in a manner similar to CASPEr but by exploiting the interaction of axions with electron spins. Ten spherical yttrium iron garnet (YIG) samples are coupled to a cylindrical copper cavity by means of an applied static magnetic field, and the resulting photon-magnon hybrid system acts as an axion-to-electromagnetic field transducer. The QUAX experiment is one of the most sensitive rf spin magnetometers ever realized, able to measure fields as small as $5.5e-19$ T with nine hours of integration time. 

The ARIADNE experiment employs an unpolarized source mass and a spin-polarized $^3$He low-temperature gas to search for a QCD-axion-mediated spin-dependent interaction: the monopole-dipole coupling. In contrast to dark matter haloscopes like CASPEr and QUAX, whose signals depend on the local dark matter density at the Earth, the signal in the ARIADNE experiment does not require axions to constitute dark matter and can be modulated in a controlled way.

\subsubsection*{Quantum Defects}
Searches for dark matter via scattering in crystals will soon run into the neutrino floor - the background of neutrinos from the sun. One path for getting beyond the neutrino floor is to develop directional detectors. Since the direction of the sun is known, the detectors can veto signals coming from the direction of the sun; dark matter interactions by contrast will results in isotropic scattering signals. One proposal for achieving this directional detection is to monitor damage tracks in crystals that occur as the scattering dark matter displaces atoms from their lattice location. These damage tracks can be measured using techniques from quantum sensing such as NV center spin spectroscopy in noncrystalline diamond. The NV center spin state is highly sensitive to the local strain in the crystal. These detectors will require a combination of imaging methods to locate and determine the direction of the damage tracks as described in \cite{Ebadi.2022} but provide a pathway towards WIMP sensitivity below the neutrino limit.

\subsection{Quantum Calorimeters}

Looking for interactions between relic dark matter with mass in the 1\,meV to 100\,MeV range and the visible sector requires the development of detectors with sensitivity to single energy depositions in the far IR (meV) to near IR (eV). Technologies that have a credible R\&D roadmap to achieve these sensitivies include but are not limited to qubits, MKIDs, TES, and SNSPDs. The precise R\&D required to improve sensitivity is sensor specific but broadly falls into three categories. First, the development of sensors from superconducting films with lower superconducting transition temperature, T$_c$, both increases the number of quasiparticles created per energy deposition (MKIDs, SNSPD) and decreases electron-phonon couplings within the film allowing for better thermal isolation (TES). Secondly, as intrinsic sensitivity of the sensor increases to dark matter, it also necessarily increases to a broad range of environmental backgrounds (blackbody IR, EMI, environmental vibrations). As such, commensurate improvements in sensor isolation from the environment must occur in parallel with these sensitivity improvements or else the noise floor will be limited by these external sources (MKIDs, TES, qubits). Third, MKIDs are currently limited by first stage amplifier noise. Implementation of lower noise temperature amplifiers or qubit inspired readout techniques.

Due to the myriad ways that dark matter could potentially interact with the visible sector (photonic, electronic and vibrations), these sensors technologies should ideally be integrated with antenna like structures and  anti-reflection coatings to maximize absorption and collection of these small excitation signals. For near IR photon collection, integration of antireflection stacks has already been shown with TES, MKIDs, and SNSPDs. For far IR collection, antennas have been integrated into TES. Finally, for Athermal phonons, Al superconducting traps have been integrated with TES. It's likely that with community effort and engagement these excitation collection and concentration technqiues can both be further refined and integrated with the remaining sensor technologies.

When operated underground in carefully designed IR tight optical cavities that have no non-instrumented insulating materials surrounded by well shielded cryostats using currently available radiopure materials and active high energy photon vetoes, low energy radioactive backgrounds should be controllable to the level of coherent nuclear scattering of solar neutrinos. The majority of experimental techniques used for light mass dark matter searches are currently limited by ``dark events", including long lived meta-stable electronic and lattice state transitions. For example, many crystalline scintillators like NaI have been considered for ERDM searches, but have afterglow, very long lived excited electronic states that produce an indistinguishable rate of single scintillation photons when they decay. 
Another example is phonon bursts due to due to multi-atom lattice reconfiguration (microfractures)  from high stress energy configurations. That are likely currently limiting all nuclear recoil light mass dark matter searches. Mitigation of these dark events requires a case-by-case understanding of the precise causal mechanism and then development of strategies to depopulate these excited states. For example, the long relaxation time scales for meta-stable electrons in excited states could be shortened by placing the crystal in an IR photon bath.  Likewise, annealing of crystalline materials has been shown to substantially decrease the lattice defect density in crystals.

\subsection{Superconducting Sensors}
\emph{Note : Contents of this section are in part taken and modified from Snowmass 2021 White Papers: Axion Dark Matter \cite{adams2022axion}, and Searches for New Particles, Dark Matter, and Gravitational Waves with SRF cavities \cite{berlin2022searches}.}

\subsubsection*{SRF Cavities}

Superconducting radio-frequency (SRF) cavities are critical components in particle accelerators. Advances in cavity performance are the results of an improved understanding of RF superconductivity and materials. In the past 50 years, new cavity processing techniques were developed to overcome limiting phenomena, such as field emission, and enhance the superconductivity. 

SRF cavities are, in their essence, extremely high quality electromagnetic resonators, devices that are now of strong active interest for quantum information science (QIS), with demonstrated record-high photon lifetime $\tau\sim 2s$ ($Q>10^{11}$) also in the quantum regime.
For quantum computing, quantum states can be stored and manipulated in electromagnetic resonators, and superconductors at milli-Kelvin temperatures are employed to sustain the coherence of the quantum states for long enough to perform complex computations. For quantum sensing, SRF cavities can furnish a large volume where very weak signals of radio-frequency photons
can be collected, with only a small fraction of photons being lost to heat at the cavity walls.


The main focus on the Superconducting Quantum Materials and Systems (SQMS) National QIS Research Center is to advance QIS through the understanding and mitigation of coherence mechanisms in 2D and 3D quantum systems, i.e. planar and cavity based, tackling the decoherence time as a primary limiting mechanisms. This SRF cavity effort is utilized also to pursue fundamental physics questions and pushing the detection sensitivity with SRF cavities. The Snowmass whitepaper \cite{berlin2022searches} summarizes opportunities to search for new particles with SRF cavities at SQMS. The focus is on dark photons and axion (or axion-like particles), either as new particles or dark matter, as well as on gravitational waves.  The search for gravity waves across the full spectrum of frequencies, particularly since their discovery by LIGO~\cite{abbott2021gwtc}, is very well motivated, potentially opening a new window onto the early Universe or new physics. In this context SRF cavities can be used to search for GW's~\cite{Berlin:2021txa}.

It is possible to explore dark photon scenarios using SRF cavities light-shining-through-wall setup. The conversion of some of the photons to
dark photons before the wall and conversion back to regular photons past the wall makes such a detection possible, if dark photons exist at a hypothesized mass and
coupling. Resonant cavities can be used on both sides of
the wall to increase the number of photons on the emitting side and to enhance the probability of conversion of
dark photons to visible ones on the receiver side. In particular, in an RF cavity the system can be designed to
search for the parametrically enhanced longitudinal coupling of the dark photon. The Dark SRF experiment at Fermilab plans to conduct such a search with ultra-high quality
cavities \cite{DarkSRF2, Raffelt:1990yz, DarkSRFpaper, graham2014parametrically}.

The following materials science and R\&D efforts are highlighted to expand current physics searches: enhance the efficiency; mitigate nonlinearitites in superconducting cavities due to TLS; reaching high-Q with the cavity in a multi-Tesla field; improving methods for frequency stability and tuning in SRF cavities. New schemes include searches with multiple cavity modes for axions or gravitational waves, nonlinear effects within the cavity walls that can mimic such a signal in particularly if the signal mode is near a harmonic; networks of SRF cavities; quantum nondemolition (QND) measurements with superconducting qubits coupled to SRF cavities.

\subsubsection*{Proposals for axion searches using SRF cavities}

The Axions and axion-like particles (ALPs) is a generalization of the QCD axion which
does not couple to QCD, but does couple to photons or
SM fermions. ALPs are well motivated in their own right in top-down constructions \cite{svrcek2006axions, arvanitaki2010string}. Like the dark photon case, Light-shining-through-wall (LSW) -type axion searches can benefit from high quality factors, which warrants the harnessing of advances in SRF technology. The necessity for a background magnetic field, however presents a challenge, as high-quality superconductivity does not survive large fields. Novel approaches to allow large magnetic fields with no degradation of Q-factor in SRF cavities are posed.

\begin{description}
\item[Two cavities with Static Field:] One technique to utilize both high-Q SRF cavities and large magnetic fields for a LSW axion search is to sequester the required magnetic fields away from the production and detection cavities \cite{janish2019axion}. With this approach neither SRF cavity is subject to large magnetic fields and neither suffers a degradation of Q-factor. However, losses in the walls of the conversion region can result in a decrease of the effective Q of the entire system.
\item[Two Cavities with a pump mode:] An alternative approach is to replace the static B-field with an oscillatory B-field, which can then be directly run inside the receiver cavity. Sources of noise due to the multi-mode setup can be mitigated by using a pump with high-Q and with the pump frequency well separated from the signal mode frequency. In addition, such noise sources can be further suppressed by optimizing the cavity geometry and material science techniques to reduce nonlinearities \cite{sikivie2010superconducting, gao2021axion}.
\item[Single-Cavity Axion Search and Euler-Heisenberg:] The EH Lagrangian makes a prediction for light-by-light scattering within the SM, which has never been observed at photon frequencies below the electron mass $m_e=511$ keV because the effect is highly suppressed at low energies.
The operating system of a  proposed experiment to search
for both the axion-induced and EH nonlinearities using
high-Q SRF cavities is described in \cite{bogorad2019probing, eriksson2004possibility, brodin2001proposal, schwinger1951gauge}. This two-cavity scheme is less sensitive to noise sources which generate nonlinearities in the pump region.


\end{description}





\subsubsection*{Qubit-based single photon counting}

The integration of a
qubit into an ultra high cavity may enable new
schemes for quantum computing and synergetically allow for employing a photon counting non-demolition measurement for DM searches. For certain DM search
schemes, it would also be beneficial to have qubits
that can operate successfully even in high magnetic
fields \cite{dixit2021searching}.

Cavity haloscopes have traditionally extracted the DM
signal via an antenna connected to a linear amplifier,
such as a Josephson Parametric Amplifier (JPA). Unfortunately, linear amplifiers contribute to their own noise power, and their minimum contribution is the standard quantum limit (SQL). SQL noise increases linearly with frequency, and thus it is necessary to subvert the SQL to make higher-mass searches feasible. 

Several ongoing or proposed experiments utilizes SRF resonators coupled to superconducting qubits to detect bosonic dark matter candidates below the SQL. Two experiments have demonstrated sub-SQL detection: HAYSTAC by implementing vacuum squeezing \cite{backes2021quantum} and SQuAD by implementing qubit-based photon counting \cite{dixit2021searching}. SQMS also plans to combine SRF cavity technology and qubit based photon counting to increase the DM search rate by several order of magnitudes. 
The Superconducting Qubit Advantage for Dark Matter (SQuAD) experiment plans to perform resonant searches for dark matter axions with DFSZ sensitivity in a broad range from 10-30 GHz using high quality factor dielectric cavities combined with
qubit-based single photon detectors which evade the quantum zero-point noise. R\&D is ongoing on developing an analogous photon counting readout based on Rydberg atoms which can be operated at the higher frequencies where qubit devices become more difficult to design and fabricate.
\newline

\subsubsection*{Networks and transduction}

Recently, it was shown that the performance of a quantum network could be utilized further to improve axion DM searches \cite{brady2022entangled}. However, the noise in the network will be incoherent among the network nodes. One can make use of distributed squeezed states to exploit the coherent nature of the DM signal. Combining quantum resources (squeezing) in a distributed-network setting can allow for a scan that is
faster by a factor of the square number of network nodes
in the ideal case. The improvement is enabled by adding
the signal at the amplitude level rather than adding powers
in the classical network case.


A quantum transduction project at Fermilab is exploring hybrid coherent resonance systems and bi-directional quantum transduction schemes to up/down- convert the microwave information to/from the optical regime and enhance the conversion efficiency at the quantum threshold, and below the SQL. Up/down photon conversion may also enable highly sensitive axion and dark photon haloscope searches in the THz regime, or microwave single photon counting in optical systems, taking advantage of optical sensing techniques, e.g.\ high precision counting in interferometry and reduced noise floor \cite{derevianko2022quantum}, also in the Snowmass LOI \emph{Opportunities for Optical Quantum Noise Reduction} \cite{derevianko2022quantumLOI}.

Transduction in the mm-wave regime is also proposed in the LOI \emph{Transduction for new Regimes in quantum sensing} \cite{Transduction_mm} as an effective way for linking the classical and quantum world, and for transporting quantum information on macroscopic scales. Low-loss mm-wave photonics could  allow preservation of quantum information at room temperature for a simpler network at laboratory scales, as well as reaching out the frequency range for axions above $\sim$10 GHz ($\sim$40 µeV) is beyond the reach of current experiments (ADMX). \newline 

\subsubsection*{Cryogenic Platform for Scaled-up Sensing Experiments}

The SQMS center at Fermilab is developing a cryogenic
platform capable of reaching millikelvin temperatures in
an experimental volume of 2 meters diameter by 1.5 meters in height \cite{hollister2021large}. The platform is designed to host a
three-dimensional qubit architecture based on SRF technology, as well as sensing experiments.

\subsubsection*{SNSPD}

Superconducting-nanowire single-photon detectors (SNSPD) are ideally suited for sensing lowcount-
rate signals due to their high internal efficiency and low dark-count rates. Recent
proposals for axion search either require SNSPDs that can operate in the presence
of large magnetic fields, or require some means of carrying the light generated by the
haloscope from the high-field region to a low-field region where the detectors can operate.
The recently established robustness of SNSPDs to operation in high fields and
their ability to operate at elevated temperatures (relative to alternative superconducting
detector technologies) make them well-suited for photon detection in the mid-infrared (meV) to visible (eV) energy range. The suitability of SNSPDs to applications requiring
low dark-count rates is illustrated by recent progress in the LAMPOST prototype search for dark photon dark-matter using these devices \cite{chiles2021first}.

\subsubsection*{Other superconducting and cryogenic sensors}

Cryogenic sensors have found a large range of applications for astroparticle detection. Due to integration complexity and thermal loading from cryogenic wiring, the ability to read out multiple detectors on a single wire with cryogenic multiplexing technologies with minimal readout noise penalty is
of utmost importance as experiments are scaled to ever larger detector counts. Several variations of SQUID multiplexers have been used to field large sensor arrays including time division multiplexing (TDM) and frequency division multiplexing (FDM) systems. One FDM implementation, the microwave SQUID
multiplexer (µmux), couples an incoming detector signal to a unique GHz-frequency resonance, thus combining the multiplexability of MKIDs with the clean separation of detection and readout interfaces. This
enables multiplexing factors up to two orders of magnitude larger than conventional cryogenic multiplexing
schemes. 

The wide frequency operation span enables large detector counts for low-bandwidth bolometric
applications such as CMB cosmology while maintaining clean interfaces between the detection and readout schemes. Additionally, the large frequency bandwidth and fast resonator response allow for cryogenic
particle detection, such as low-mass threshold dark matter searches, beta decay end point measurements
to determine the lightest neutrino mass, and coherent elastic neutrino-nucleon scattering.

The CUPID collaboration in the snowmass whitepaper \emph{Toward CUPID-1T}~\cite{armatol2022toward} presents a  series of projects underway that will provide advancements in background reduction, cryogenic readout, and physics searches, all moving toward the next-to-next generation CUPID-1T detector.
Neutron-transmutation doped thermistors (NTDs) are expected as part of the baseline design for CUPID. Multiple modes of superconducting sensors are under development as we look toward
CUPID-1T: Microwave Kinetic Inductance Detectors (MKIDs), Metallic Magnetic Calorimeters (MMCs),
and high- and low-impedance Transition Edge Sensors (TESes). 

\section{Common Areas for Development}
\begin{itemize}
    \item Back action evasion:
    Back action evading schemes and squeezing techniques can enhance the sensitivity measurement of quantum sensors down to the SQL. Many experiments (for example NMR experiments to axion dark matter) will have their sensitivity limited by quantum back action and techniques will need to be developed for experiments approaching fundamental projection noise sensitivity limits. One of the purposes of new transduction projects is to leverage both microwave and optical sensing techniques as a way means to implement back action evasion. 
  
    \item Supporting technologies (material science, laser, cavities, magnets, etc.):
    Several sensing experiments are enabled by SRF cavities with high-Q. Material studies, efforts toward mitigating TLS-driven losses, and enhancing operation under multi-Tesla magnetic field can provide new resources for quantum sensors. Collaboration with non-HEP groups such as those that drive quantum computing focused material science studies may stimulate interest in developing loss mitigation strategies that further improve SRF cavity Q. Similarly, many experiments rely on the use of high-field magnets. Efforts to increase the magnitude, uniformity and scale (larger magnet bores) can result in direct improvements to the experimentally reachable parameter space.
    
    \item Infrastructures: The same characteristics that allow quantum sensors to probe new parameter space allows them to be sensitive to a wide range of noise sources. In some cases, experiments may need to be placed in underground labs to avoid noise sources such as cosmic rays or maintain radio purity of sensor materials. Development of shared infrastructure, e.g.\ underground facilities with cryogenic and or magnetic capabilities could enable the advancement of multiple experimental techniques in a single facility.  
    
    \item SBIR program, interaction with companies: 
    DOE programs for commercialization and technology transfer programs, such as SBIR/STTR, provides platforms and resources to develop technology for quantum sensors and HEP. With the rapid rise of commercial sector quantum computing  and associated technologies, new opportunities are emerging for interactions between government sponsored researchers and the commercial sector.

\end{itemize}

\bibliography{main}

\clearpage

\end{document}